# Multiferroicity with high-*Tc* in ceramics of the YBaCuFeO$_5$ ordered perovskite

B. Kundys,[‡] A. Maignan, and Ch. Simon

*Laboratoire CRISMAT, UMR 6508 CNRS/ENSICAEN, 6 bd du Maréchal Juin, F-14050 CAEN Cedex 4, Normandy 14050, France*

A dielectric anomaly has been found near the incommensurate to commensurate antiferromagnetic phase transition ($T_{N2} \approx 230$ K) in YBaCuFeO$_5$ ceramics, a compound which crystallizes in an ordered perovskite structure. The existence of electric polarization below $T_{N2}$ suggests the magnetism induced charge polarization effect that is also confirmed by its strong magnetic field dependence below $T_{N2}$. Accordingly, the peak near $T_{N2}$ of the magnetodielectric effect indicates a maximum of magnetodielectric susceptibility near the spin reorientation transition. Considering the abundance of magnetic compounds which structures derive from the crystal perovskite, these results might open up the way toward the control of electric polarization near room temperature.

The search for multiferroic materials is motivated by the need of nonvolatile random access memories for which the electric polarization (magnetization) is controlled by magnetic field (electric field) or vice versa. The use of such materials would be extremely beneficial for next generation electronics for continuing the ongoing miniaturization process as well as optimization of reading/writing speed and power consumption. For bulk ceramics, combining antiferromagnetism and electric polarization is a tricky problem as such properties are exclusive in most materials.[1,2] In that respect, the discovery of magnetization induced ferroelectricity in noncollinear (spiral) antiferromagnet (AF) (Ref. [3] and [4]) for which the inversion symmetry is broken, has opened new possibilities. As spiral antiferromagnetic structures are usually resulting from competing magnetic interactions – triangular frustrated magnetic networks—the temperatures of magnetoelectric coupling are often limited by the too low magnetic ordering temperatures.[3,5–11] However, crystals of CuO (Ref. [12]) tenorite oxide have been recently shown to be a remarkable first exception with spiral induced electric polarization in the 213–230K range where an incommensurate antiferromagnetic structure is observed. This report has motivated us to look backward to other incommensurate antiferromagnetic oxides characterized by higher magnetic ordering temperatures than those found in typical frustrated systems. As in a majority of copper oxides with structures deriving from the perovskite characterized by magnetic square lattice, the copper mixed-valency tends to make them too conducting. We looked for compounds of similar structures but with mixed transition-metal cations at the metal (*M*) site. Accordingly, the oxygen deficient ordered perovskites "112" structures of generic formula *L*Ba*M*$_2$O$_5$ [*L*=Y or a lanthanide and *M*=(Cu/Fe) (Ref. [13]), Co (Ref. [14]), Mn (Ref. [15])] are attractive candidates. Their structures can be described by starting from the *L*Ba*M*$_2$O$_6$ double perovskites with a layer ordering of the $L^{3+}$ and $Ba^{2+}$ cations and then by removing the oxygen from the (LO) plane. This leads to the existence of double layers of square pyramids [*M*O$_5$]$_2$ linked by their apical oxygen (Fig. 1(a)). In addition to the ordering at the (Ln,Ba) and oxygen sites, a third layered ordering mechanism could theoretically be created in the bipyramids by using two different *M'* and *M"* cations leading to the *L*Ba*M'M"*O$_5$ formula with ordered layers of *M'*O$_5$ and *M"*O$_5$ square pyramids (Fig. 1(b)). The difference between disordered and ordered *M'*O$_5$ and *M"*O$_5$ layers in bipyramids is a symmetry change from the centrosymmetric *P*4/*mmm* tetragonal space group for the disordered phase to a noncentrosymmetric *P*4*mm* one in the ordered phase.[16] The YBaCuFeO$_5$ compound, first discovered in Caen,[13] belongs to this structural type with conflicting reports about the Cu$^{2+}$/Fe$^{3+}$ ordering in the (*M*$_2$O$_{10}$) bilayers.[16–18] Interestingly, its commensurate AF structures ($T_{N1} \approx 440$K) rearranges in an incommensurate spiral structure below $T_{N2} \approx 230$ K. In the present paper, we report on the evidence for electric polarization induced by incommensurate AF in YBaCuFeO$_5$. The YBaCuFeO$_5$ is a second magnetoelectric oxide after CuO in which electric order appears simultaneously with noncollinear antiferromagnetic order that exhibits a Curie temperature overpassing 200 K. Moreover, considering time consuming single crystals growth and orientation, these results obtained on ceramic sample suggest that search for new multiferroic materials may be significantly speed up focusing on polycrystalline magnetic multiferroic materials. The ceramic samples were prepared by conventional solid-state reaction according to the synthesis conditions reported in Ref. 16. Bars obtained by pressing that powder were then reacted at 1000 °C for 24 h and then quenched in ambient air. The room temperature x-ray powder diffraction (XRPD) of the reacted materials confirm the YBaCuFeO$_5$ tetragonal structure (from XRPD, it is not possible to distinguish P4/*mmm* from P4*mm*) in good agreement with existing data for that phase. Magnetodielectric and polarization measurements were carried out in a PPMS Quantum Design cryostat. The capacitance and dielectric loss were measured using Agilent 4248A RLC bridge at different frequencies. Silver paste was used to make electrical contacts to the sample. The polarization was measured with a Keithley 6517A electrometer. These temperature dependent measurements are given in Fig. 2 at zero applied magnetic field together with the magnetization curve. To measure electric polarization, the sample was first cooled in electric field (270 kV/m) from 270 down to 10 K and then the electric field was removed and P(T) curve was recorded using a temperature sweep of 20 K/min heating rate. Before heating from 10 to 300 K a time dependence of electric charge was recorded for 2 ks to ensure polarization stability. The magnetization versus T curve collected in 0.3 T after a zero-field-cooling process reproduces also the data of the literature with the characteristic transition showing its maximum near (225 K (Fig. 2(a)). This value corresponds to $T_{N2}$ at the AF transition from incommensurate below $T_{N2}$ to commensurate for $T_{N2}<T<T_{N1}$.[16] A dielectric anomaly (Fig. 2(b)) starts at about 247 K, i.e., about 20 K above $T_{N2}$. The temperature position of the dielectric anomaly has no frequency dependence confirming its intrinsic origin (Fig. 2(b)). The dielectric anomaly is also correlated with dielectric loss of the sample (inset of Fig. 2(b)). However the magnitude of dielectric anomaly decreases as frequency increases confirming a frequency dependence of the dielectric permittivity similarly to that observed in CuO.[12] The existence of an electric polarization (Fig. 2(b)) related to the dielectric anomalies at $T_{N2}$ implies spin orientation induced magnetoelectric coupling in the present polycrystalline sample. The electric polarization can be also switched with electric field using different electric field cooling procedures (Fig. 2(c)). The development of the electric polarization (Fig. 2(c)) with an inflection point near $T_{N2}$, shows that its origin is related to the magnetic transition (Fig. 2(a)). To prove it, the electric polarization under different magnetic fields H has been recorded (Fig. 3). To measure the electric polarization under magnetic field H, an external magnetic field was applied at 270 K, then an electric field of 270 kV/m and finally, the sample was cooled down to 10 K. As it can be observed in Fig. 3, the H application makes the electric polarization P decreasing and the transition temperature shifting toward lower temperatures. Similar behavior has been observed in other compounds[5] and is linked to a change by externally applied magnetic field of the spin configuration responsible for electric polarization. The measurements of the dielectric permittivity as a function of magnetic field reveal symmetrical branches for positive and negative applied magnetic field with a hysteresis presenting a butterfly shape for temperatures in vicinity of $T_{N2}$ (only 4 curves over 11 are shown in Fig. 4). Although the shape of the dielectric permittivity versus magnetic field remains similar for electrically polar region in the incommensurate AF structure ($T<T_{N2}$), the magnitude of the magnetodielectric effect is strongly $T$ dependent. As it can be observed from the selected curves of Fig. 4, as the temperature increases, the magnitude of the magnetodielectric effect increases and it reaches a maximum for $T$=170 K. As the temperature approaches the characteristic temperature $T_{N2} \approx 230$K of the commensurate-incommensurate AF transition, the magnitude of the relative magnetodielectric effect decreases and practically no magnetodielectric effect is observed for $T=T_{N2}$. This demonstrates that in the commensurate AF state ($T>T_{N1}$=440K), the magnetodielectric effect tends to be suppressed so that a maximum of magnetodielectric effect is observed before the spin reorientation transition (Fig. 5). A similar behavior, with a magnetodielectric effect which increases as temperature approaches the transition as shown by Fig. 5, was already observed for Cr$_2$O$_3$ (Ref. 19) and seems to be a general feature for many antiferromagnetic magnetoelectric samples if one takes into account the phenomenological model analysis.[20] It has to be noted that attempts to switch electric polarization by 180° with electric fields at different temperatures were unsuccessful. Thus, we assume that our compound either just electrically polar of electric coercive force is too high.

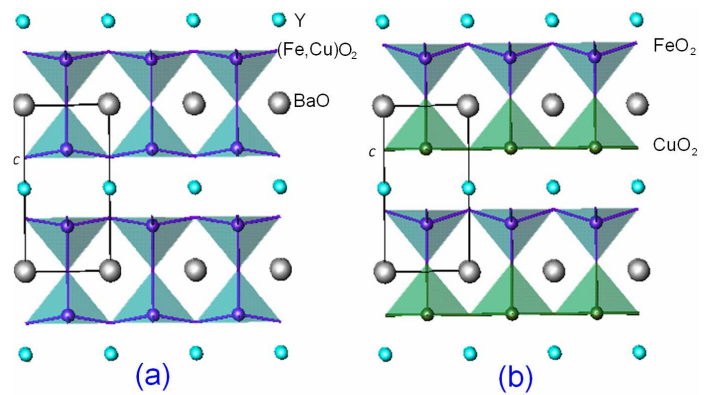

FIG. 1. (Color online) (a) Schematic drawing of the *L*Ba*M*$_2$O$_5$ structure for *L*=Y$^{3+}$ and *M*=Cu$_{2+}$, Fe$_{3+}$. )b) In the *P*4*mm* acentric structure, the different positions of Fe and Cu in the pyramids might be favorable to electric polarization along the *c*-axis.

---

[‡]Electronic mail: kundys@.gmail.com (Bohdan Kundys).

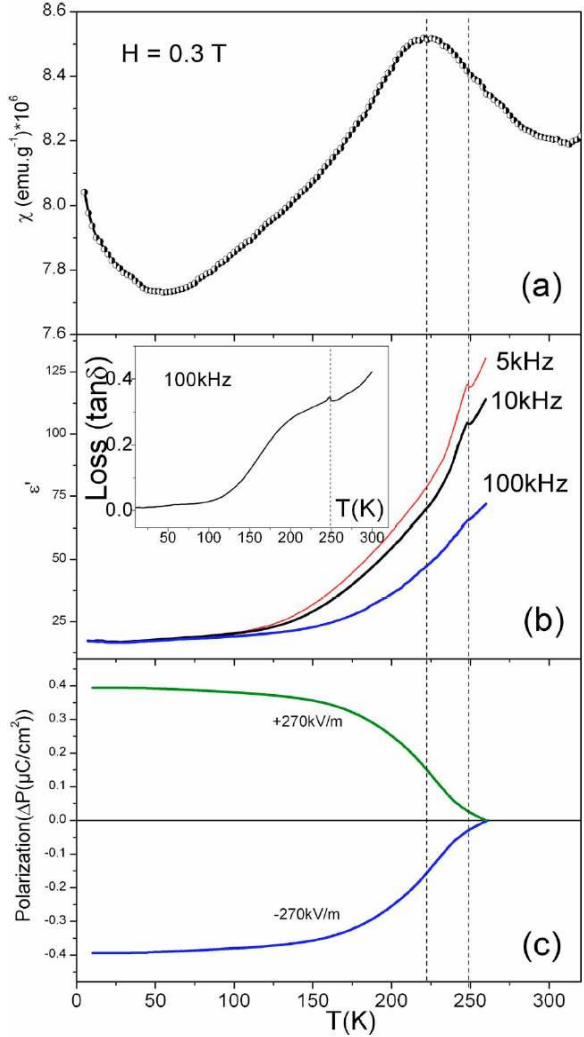

FIG. 2. (Color online) ZFC (0.3T) magnetization, (a) dielectric permittivity measured under heating at different frequencies, (b) and electric polarization (c( of the YBaCuFeO$_5$ ceramics as a function of temperature. Inset of figure (b) shows dielectric loss at 100 kHz on heating.

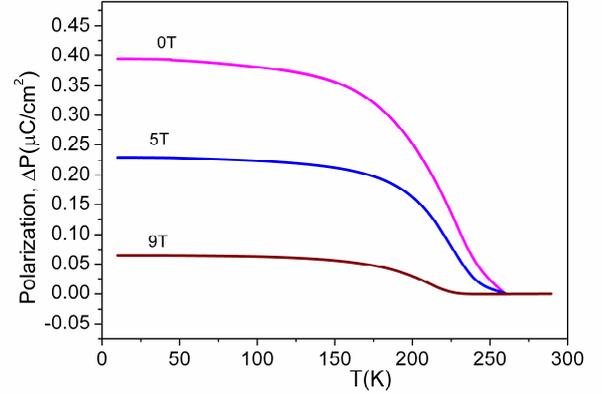

FIG. 3. (Color online) The electric polarization of YBaCuFeO$_5$ ceramics recorded in electric after 270 kV/m cooling procedure with different magnetic fields.

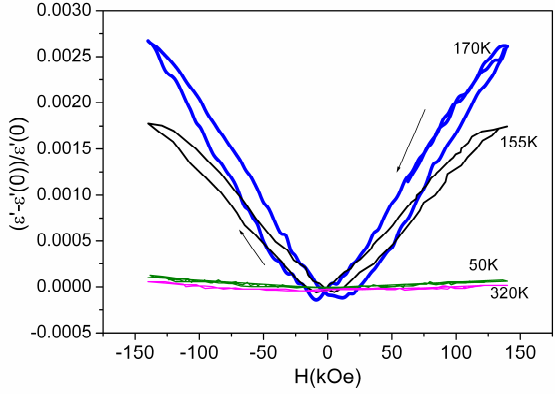

FIG. 4. (Color online) Relative dielectric permittivity at 100 kHz as a function of magnetic field taken at different temperatures for YBaCuFeO$_5$. The arrows indicate increasing and decreasing magnetic fields.

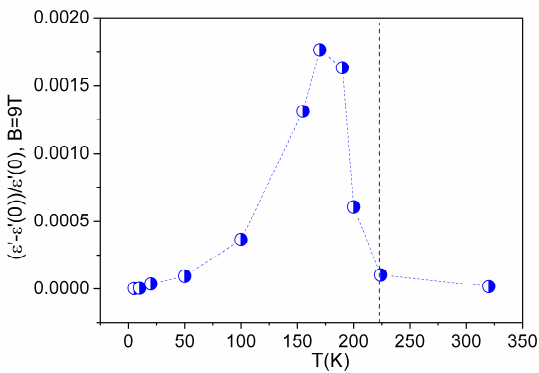

FIG. 5. (Color online) Temperature dependence of the relative dielectric permittivity at 100 kHz taken at 9 T for YBaCuFeO$_5$.

The evidence for this multiferroic (i.e., coupled magnetic and electric orders) behavior of YBaCuFeO$_5$ below $T_{N2} \approx 230$K has to be connected to the incommensurate AF structure of YBaCuFeO$_5$. In that structure, the AF structure sets only at short range[16] which might reflect some degree of disorder between the magnetic interactions of the Fe$^{3+}$ ($S$=5/2) and/or Cu$^{2+}$ ($S$=1/2) cations either in the [CuFeO$_{10}$] blocks of bipyramids or between the successive blocks. As below $T_{N2}$, the magnetic structure refinement is improved by using a $P4mm$ acentric space group,[18] the removal of the mirror plane perpendicular to the fourfold axis as compared to a $P4/mmm$ space group, could explain the existence of an electric polarization along the fourfold axis, i.e., along the $c$ direction. Such an explanation was first proposed in the case of CuO for which the mirror plane of the nonpolar $2/m$ space group removed by spiral magnetic order leads to the polar space group 2.[12] In YBaCuFeO$_5$, this effect would have to be related to the existence of two different positions for Cu$_{2+}$ and Fe$_{3+}$ in square pyramids (Fig. 1(b)), the former lying closer to the basal plane than the latter which might create dipoles along the fourfold axis of the pyramids.

In conclusion, it turns out that the square lattice in perovskites structures showing some layering perpendicular to the fourfold axis, induced by the Y/Ba and oxygen deficiency in the present structure, favors complex, and disordered AF structures which may generate induced electric polarization along the stacking direction. In this structural type, the mixed occupation at the M site (in the $L$Ba$M'M''$O$_5$ structures of these ordered perovskites) generates some topological disorder and also the $M$O$_5$ pyramidal coordination strongly affects the hybridization of the $e_g$ orbitals and O$_{2p}$ ligands.[21] Both factors are most probably important to explain the magnetoelectric coupling in the present YBaCuFeO$_5$ ceramic.